\documentclass[%
 reprint,
superscriptaddress,
 amsmath,amssymb,
 aps,
]{revtex4-1}

\usepackage{graphicx}
\usepackage{dcolumn}
\usepackage{bm}
\usepackage{xcolor}

\begin{document}

\preprint{APS/123-QED}

\title{Spin interactions and magnetic order in the iron oxychalcogenides BaFe$_{2}$$Q_{2}$O}

\author{B. D. Coles}
\affiliation{School of Physical Sciences, University of Kent, Canterbury, Kent, CT2 7NH, U.K.}
\author{A. D. Hillier}
\affiliation{ISIS Facility, Rutherford Appleton Labs, Chilton Didcot, OX11 0QX, U.K.}
\author{F. C. Coomer}
\affiliation{ISIS Facility, Rutherford Appleton Labs, Chilton Didcot, OX11 0QX, U.K.}
\author{N. C. Bristowe}
\affiliation{School of Physical Sciences, University of Kent, Canterbury, Kent, CT2 7NH, U.K.}
\author{S. Ramos}
\thanks{S.Ramos-Perez@kent.ac.uk}%
\affiliation{School of Physical Sciences, University of Kent, Canterbury, Kent, CT2 7NH, U.K.}
\author{E. E. McCabe}
\thanks{e.e.mccabe@kent.ac.uk}%
\affiliation{School of Physical Sciences, University of Kent, Canterbury, Kent, CT2 7NH, U.K.}

\date{\today}

\begin{abstract}
The ability to tune the iron chalcogenides BaFe$_{2}Q_{3}$ from Mott insulators, to metals and then superconductors with applied pressure has renewed interest in low-dimensional iron chalcogenides and oxychalcogenides. We report here a combined experimental and theoretical study on the iron oxychalcogenides BaFe$_{2}Q_{2}$O ($Q$=S, Se) and show that their magnetic behaviour results from nearest-neighbour magnetic exchange interactions via oxide and selenide anions of similar strength, with properties consistent with more localised electronic structures than those of  BaFe$_{2}Q_{3}$ systems.



\end{abstract}

\pacs{Valid PACS appear here}
\maketitle

\maketitle
\section{Introduction}

The study of iron oxychalcogenide materials has developed in parallel to the effort to understand iron based superconductivity ~\cite{Kamihara_JACS}. The strength of the correlations between electrons and the proximity of these materials to a Mott insulator transition are fundamental questions about the iron pnictide and iron chacogenide superconductors ~\cite{Si_Abrahams, Norman} that need to be understood in these materials but also in related systems that do not display superconductivity. In this context, iron oxychalcogenides such as $X_{2}$O$_{2}$Fe$_{2}$O$Q_{2}$ (with $X$= La, Nd and $Q$ = S, Se) are an important family to investigate. Firstly because they are structurally related to the superconductors and secondly because substitution of the chalcogenide ion allows for some tuning of the electronic bandwidth (and hence their proximity of a metal-insulator transition) ~\cite{band_narrowing}. 

The work presented in this manuscript was prompted by the report of superconductivity in the 2-leg ladder materials BaFe$_{2}Q_{3}$ ($Q$ = S, Se) under pressure~\cite{Yamauchi,Takahashi,Ying}. This discovery has highlighted the importance of understanding not only the role of electronic structure and degree of electronic correlation in iron based superconductors, but also that of the crystal structure and dimensionality~\cite{Hisada,Pizarro,Zhang,Gu}.

The crystal structure of BaFe$_{2}Q_{3}$ materials is composed of double-chains of edge-linked Fe$Q_{4}$ tetrahedra (the 2-leg ladders) with ladders well separated from one another (by $\sim$6 \AA) by barium cations~\cite{Caron,Steinfink}. This results in much stronger intra-ladder interactions than the interactions between ladders,~\cite{Mourigal,Sefat}, with experimental work suggesting short-ranged antiferromagnetic (AFM) correlations well above the Neel temperature $T_{\mathrm{N}}$~\cite{Lei2,Caron}. Given the ratio of intraladder:interladder exchange, theoretical work supports the description of these systems as "pseudo-1D ladders", highlighting the nearly one-dimensional nature of the exchange interactions~\cite{Sefat}. At ambient pressure, they can be described as orbital-selective Mott insulators~\cite{Yamauchi,Mourigal}. With increased pressure, BaFe$_{2}Q_{3}$ undergoes first an insulator-metal phase transition, and at higher pressures, a transition to a superconducting state~\cite{Yamauchi,Takahashi,Ying}. These remarkable materials have given renewed interest in iron ladder compounds~\cite{Zhang,Lai} and specifically, how their magnetic and electronic structures compare with related systems in terms of dimensionality and electron correlation. 

The Mott insulating oxychacogenides BaFe$_{2}Q_{2}$O studied in this work share some common features with the 2-leg ladder $A$Fe$_{2}Q_{3}$ systems. Both contain tetrahedrally-coordinated Fe$^{2+}$ cations but in BaFe$_{2}Q_{2}$O, the Fe$Q_{3}$O tetrahedra are corner-linked via oxide anions (along [010]) forming the "rungs" of ladders, and corner-linked via chalcogenide anions (along [100]) forming the lengths of the ladders. These ladders are linked across edges of the Fe$Q_{3}$O tetrahedra to give buckled Fe-$Q$-O layers, separated by barium cations (Figure \ref{BaFe2Q2O_structure_susceptibility}$a$). The magnetic exchange interactions in BaFe$_{2}Q_{2}$O materials have been suggested to be quite anisotropic, with AFM Fe - O - Fe $J_{1}$ exchange along the ladder rungs thought to dominate~\cite{Han,Popovic,Lei}. This has given rise to their description as "spin ladders"~\cite{Popovic,Valldor,Huh,Guo}, prompting our investigation using neutron powder diffraction (NPD) and muon spin relaxation methods to investigate their magnetic behaviour. In contrast to $A$Fe$_{2}Q_{3}$ systems, the Fe$^{2+}$ coordination environment in BaFe$_{2}Q_{2}$O materials (with coordination by both oxide and softer chalcogenide anions, and in buckled layers) is thought to narrow the Fe 3d bands~\cite{Han} and so comparison with $A$Fe$_{2}Q_{3}$ systems gives some insight into the effect of band narrowing in these materials. Our experimental work is complemented by a theoretical study to investigate the magnetic exchange interactions as a function of on-site Coulomb potential $U_{Fe}$; this illustrates how the BaFe$_{2}Q_{2}$O materials differ from the spin-ladder $A$Fe$_{2}Q_{3}$ systems and reveals the source of magnetic frustration and spin dynamics suggested by other experimental studies.

\section{Methods}
Polycrystalline samples of BaFe$_{2}$S$_{2}$O and BaFe$_{2}$Se$_{2}$O were prepared by the solid state reaction of stoichiometric quantities of BaO (99.99\%), Fe powder (99+\%) and S powder (99.5\%) or Se powder (99.5+\%). The reagents were weighed and ground by hand in an agate pestle and mortar in an Ar filled glove box (H$_{2}$O $<$ 0.5 ppm, O2 $<$ 0.5 ppm) and placed in small alumina crucibles. These were placed inside quartz reaction tubes which were evacuated and sealed under vacuum. The reaction tubes were heated slowly to a reaction temperature of 800$^\circ$C for BaFe$_{2}$S$_{2}$O and to 740$^\circ$C for BaFe$_{2}$Se$_{2}$O, held at this temperature for 24 hours and allowed to cool in the furnace. The reaction mixtures were then reground, pelletized and sealed again in evacuated quartz tubes and heated slowly to the reaction temperature for a further 24 hours before cooling in the furnace.
Initial characterisation was carried out using a Rigaku Miniflex600 X-ray powder diffractometer with copper source and nickel filter. Field-cooled (FC) and zero-field-cooled (ZFC) magnetic susceptibility data were collected on warming (at a rate of 5$^{\circ}$ C min$^{-1}$) for $\sim$ 0.1 g of BaFe$_{2}$S$_{2}$O and on $\sim$ 0.04 g of BaFe$_{2}$Se$_{2}$O in fields from 1000 - 60000 Oe (see section IIIA). Neutron powder diffraction (NPD) data were collected for BaFe$_{2}$S$_{2}$O on the high-flux diffractometer D20 at the Institut Laue Langevin (Grenoble, France) with a neutron wavelength of 2.41 \AA. The powder was placed in a 10 mm diameter cylindrical vanadium can (to a height of 2.5 cm) and data were collected from 5-130$^{\circ}$ 2$\theta$. Four 10 minute scans were collected at 1.8 K and 10 minute scans were collected on warming at 2 K min$^{-1}$ to 290 K. NPD data were collected for BaFe$_{2}$Se$_{2}$O on the time-of-flight (TOF) diffractometer Wish on target station 2 at the ISIS spallation neutron and muon source (Rutherford Appleton Laboratory, U.K.). The powder was placed in a 6 mm diameter cylindrical vanadium can (to a height of 1 cm). A 60 minute (40 $\mu$Amp) scan was collected at 2 K before the sample was heated to 245 K, with 60 minute scans collected at 80 K and 180 K, and 20 minute (13 $\mu$Amp) scans collected at intermediate temperatures at 5 K intervals. Rietveld refinements ~\cite{Rietveld} were performed using TopasAcademic software~\cite{Coelho2003, Coelho2012}. For refinements using constant wavelength NPD data, the diffractometer zero point and neutron wavelength were refined using data collected at 160 K for which lattice parameters were known from XRPD analysis and were then fixed for subsequent refinements. A background was refined for each refinement, as well as unit cell parameters, atomic positions and a pseudo-Voigt peak shape. Constant-wavelength NPD data were of lower resolution and only data up to 70$^{\circ}$ 2$\theta$ were used in refinements, therefore a single global isotropic temperature factor was used for all sites.

Temperature dependent muon spin relaxation data in zero applied field were collected at EMU (ISIS spallation neutron and muon source, Rutherford Appleton Laboratory, U.K.). A closed cycle refrigerator was used to vary the temperature between 400 K and 10 K.  The sample was 0.683 g of material in powder form, contained within Ag foil pouches (1.5 cm$^{2}$). 

First principles calculations based on Density Functional Theory (DFT) were employed to determine the spin exchange interactions. All simulations made use of the VASP package~\cite{Kresse1993,Kresse1996}, version 5.4.1. We chose the PBEsol~\cite{Perdew2008}+$U$ exchange correlation potential within the Liechtenstein~\cite{Liechtenstein1995} framework, where the effective on-site Coulomb and exchange parameters, $U$ and $J$, were varied on the Fe $d$ electrons within a sensible range of values~\cite{Han}. To converge the total energy, force and stress to within 0.5 meV/u.c., 0.5 meV/\AA~and 0.02 GPa respectively, we found that an 800 eV plane wave cutoff and 6$\times$2$\times$4 $k$-point mesh for the 12 atom unit cell were necessary. Tests were made to check that the energy difference between single 12 atom and doubled 24 atom supercells were kept below 0.5 meV/u.c. using these parameters. Projector Augmented Wave pseudopotentials~\cite{Bloechl1994} were used in the calculations with the following valence electron configuration: 5$s^2$ 5$p^6$ 6$s^2$ (Ba), 3$p^6$ 4$s^2$ 3$d^6$ (Fe), 4$s^2$ 4$p^4$ (Se), 3$s^2$ 3$p^4$ (S) and 2$s^2$ 2$p^4$ (O). Atomic coordinates and lattice vectors were frozen to the low temperature neutron data for the BaFe$_{2}$S$_{2}$O and BaFe$_{2}$Se$_{2}$O systems.

\section{Results}
\subsection{Magnetic susceptibility}
\begin{figure}[t] 
\includegraphics[width=0.95\linewidth]{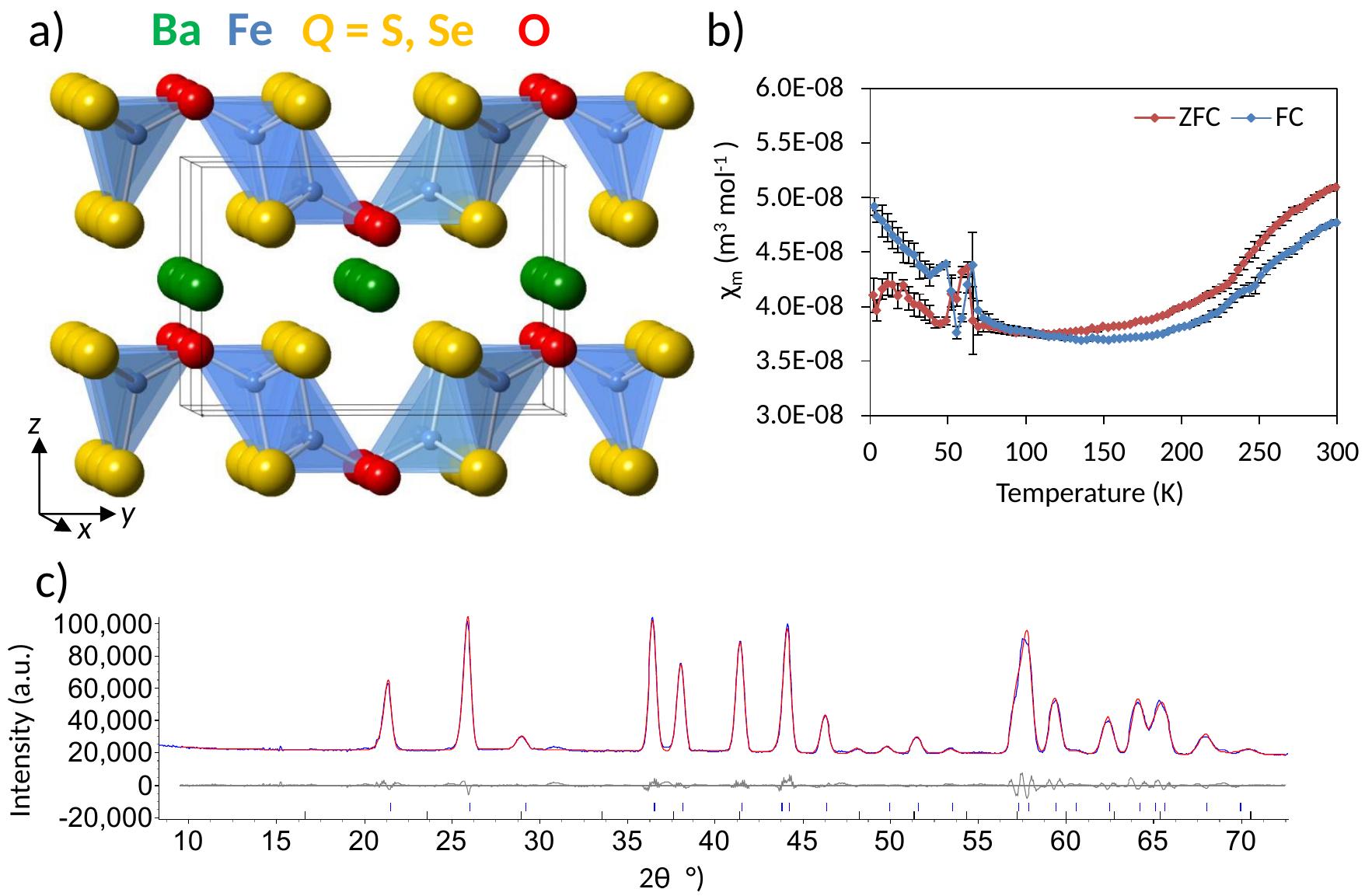}
\caption{[color online]  a) nuclear cell of BaFe$_{2}Q_{2}$O showing Ba, Fe, $Q$ and O ions in green, blue, yellow and red, respectively b) shows magnetic susceptibility data for BaFe$_{2}$S$_{2}$O and c) shows Rietveld refinement profiles for BaFe$_{2}$S$_{2}$O using 293 K NPD data. Observed, calculated and difference profiles are shown in blue, red and grey, respectively; upper blue ticks and lower black ticks show reflection positions for the BaFe$_{2}$S$_{2}$O and for Fe$_{3}$O$_{4}$ (1.13(2)\% by weight), respectively.}
\label{BaFe2Q2O_structure_susceptibility}
\end{figure}

Initial measurements for BaFe$_{2}$Se$_{2}$O in 1000 Oe applied field (supplementary material) suggested three phase transitions, consistent with single crystal measurements reported by Lei \textit{et al.}~\cite{Lei}. However, magnetisation measurements as a function of field at 300 K indicated the presence of a FM component that saturates in a field of 10000 Oe, and analogous results were found for BaFe$_{2}$S$_{2}$O (supplementary material). Field-cooled and zero-field cooled susceptibility data (Figure \ref{BaFe2Q2O_structure_susceptibility}$b$) were obtained by subtracting data collected in applied field of 45000 Oe from those collected at 55000 Oe (above the saturation level of the FM impurity). This method was used to subtract the contribution from ferromagnetic impurities (that order within this temperature range) and reveal the behaviour of the bulk sample. These results indicate that Curie-Weiss behaviour is not observed over the whole temperature range for either sample, and the changes in slope at 240 K for BaFe$_{2}$Se$_{2}$O and at 250 K for BaFe$_{2}$S$_{2}$O indicate the development of long-range magnetic order below these temperatures, consistent with other reports~\cite{Popovic, Lei, Valldor}. In these corrected data, the anomaly in susceptibility at $\sim$115 K, also observed by Lei \textit{et al.}~\cite{Lei}, is absent, suggesting that this may arise from a FM impurity phase such as Fe$_{3}$O$_{4}$~\cite{Verwey}. However, the low temperature feature (at $T_{\mathrm{2}}\sim$59 K for BaFe$_{2}$S$_{2}$O; at $T_{\mathrm{2}}\sim$40 K for BaFe$_{2}$Se$_{2}$O) is still observed and may indicate freezing of some spin dynamics, as discussed further below.

\subsection{Room temperature NPD data}
NPD data collected above $T_{\mathrm{N}}$ for both BaFe$_{2}$Se$_{2}$O and BaFe$_{2}$S$_{2}$O are consistent with the reported crystal structures~\cite{Han, Valldor}. Preliminary refinements were carried out to investigate sample stoichiometry, with the Ba site occupancy fixed at unity and a single overall temperature factor. These refinements indicated that both samples were close to their ideal stoichiometries although slightly iron- and chalcogenide-deficient (BaFe$_{1.966(3)}$Se$_{1.982(3)}$O$_{0.991(3)}$ and BaFe$_{1.935(5)}$S$_{1.87(1)}$O$_{0.985(8)}$); sites were assumed to be fully occupied in subsequent analysis of the long-range magnetic structures. Refinements to investigate possible anion-disorder indicated full ordering of oxide and chalcogenide anions for both samples. Final refinement details and profiles are given in Table \ref{table_refine_rt} and Figure \ref{BaFe2Q2O_structure_susceptibility}$c$. Both samples were found to contain traces of impurities including Fe$_{3}$O$_{4}$, consistent with magnetic susceptibility data discussed above (BaFe$_{2}$Se$_{2}$O contained 3.66(3)\% Fe$_{3}$O$_{4}$ by weight, and 2.02(1)\% FeSe by weight; BaFe$_{2}$S$_{2}$O contained 1.13(1)\% Fe$_{3}$O$_{4}$ by weight). There was no evidence in these room temperature NPD data of any diffuse scatter that might result from short-range magnetic correlations or short-range order.

\begin{table}[ht]
\caption{Refinement details and selected distances, bond lengths and angles from Rietveld refinements using 293 K NPD data for BaFe$_{2}$S$_{2}$O and 275 K NPD data for BaFe$_{2}$Se$_{2}$O using $Pmmn$ nuclear model for both.}
\centering
\begin{tabular} {c c c}
\hline\hline
X & $Q$= S, 300 K &  $Q$= Se, 275 K\\
\hline\hline
$a$ (\AA)& 4.0038(2) & 4.13425(9) \\
$b$ (\AA) & 9.5729(6) & 9.8516(1) \\
$c$ (\AA) & 6.4765(4) & 6.7202(1) \\
volume (\AA $^3$) & 248.23(2) & 273.705(7) \\
Ba 2$a$ $z$ & 0.5276(6) & 0.5103(2) \\
Ba $U_{\mathrm{iso}}$ $\times 100$ (\AA $^2$) & 1.4(2) & 1.07(5) \\
Fe 4$e$ $y$ & 0.6684(2) & 0.6642(6) \\
Fe 4$e$ $z$ & 0.8797(3) & 0.87971(9) \\
Fe $U_{\mathrm{iso}}$ $\times 100$ 100 (\AA $^2$) & 1.4(2) & 1.74(3) \\
Se 4$e$ $y$ & 0.7880(8) & 0.79261(7) \\
Se 4$e$ $z$ & 0.766(1) & 0.7588(1) \\
Se $U_{\mathrm{iso}}$ $\times 100$ 100 (\AA $^2$) & 1.4(2) & 1.39(3) \\
O 2$b$ $z$ & 0.7310(7) & 0.7394(1) \\
O $U_{\mathrm{iso}}$ $\times 100$ 100 (\AA $^2$) & 1.4(2) & 1.49(5) \\
$R_{wp}$  (\%) & 3.40 & 3.68 \\
$R_{p}$  (\%) & 2.51 & 4.13 \\
$\chi^{2}$  & 33.32 & 13.75 \\
Fe - Fe [010] (\AA) & 3.225(5) & 3.235(1) \\
Fe - Fe [111] (\AA) & 2.979(3) & 3.1217(8) \\
Fe - O (\AA) & 1.878(3) & 1.8723(8) \\
Fe - $Q$ [001] (\AA) & 2.331(7) & 2.4660(9) \\
Fe - $Q$ [110] (\AA) & 2.420(5) & 2.5560(5) \\
Fe - O - Fe ($^{\circ}$) & 118.3(3) & 119.54(7) \\
Fe - $Q$ - Fe [100] ($^{\circ}$) & 111.6(3) & 107.95(3) \\
Fe - $Q$ - Fe [111] ($^{\circ}$) & 77.63(1) & 76.84(2) \\
\hline
\label{table_refine_rt}
\end{tabular}
\end{table}

\subsection{Low temperature NPD analysis and magnetic structure}
\begin{figure}[t]
\includegraphics[width=0.95\linewidth]{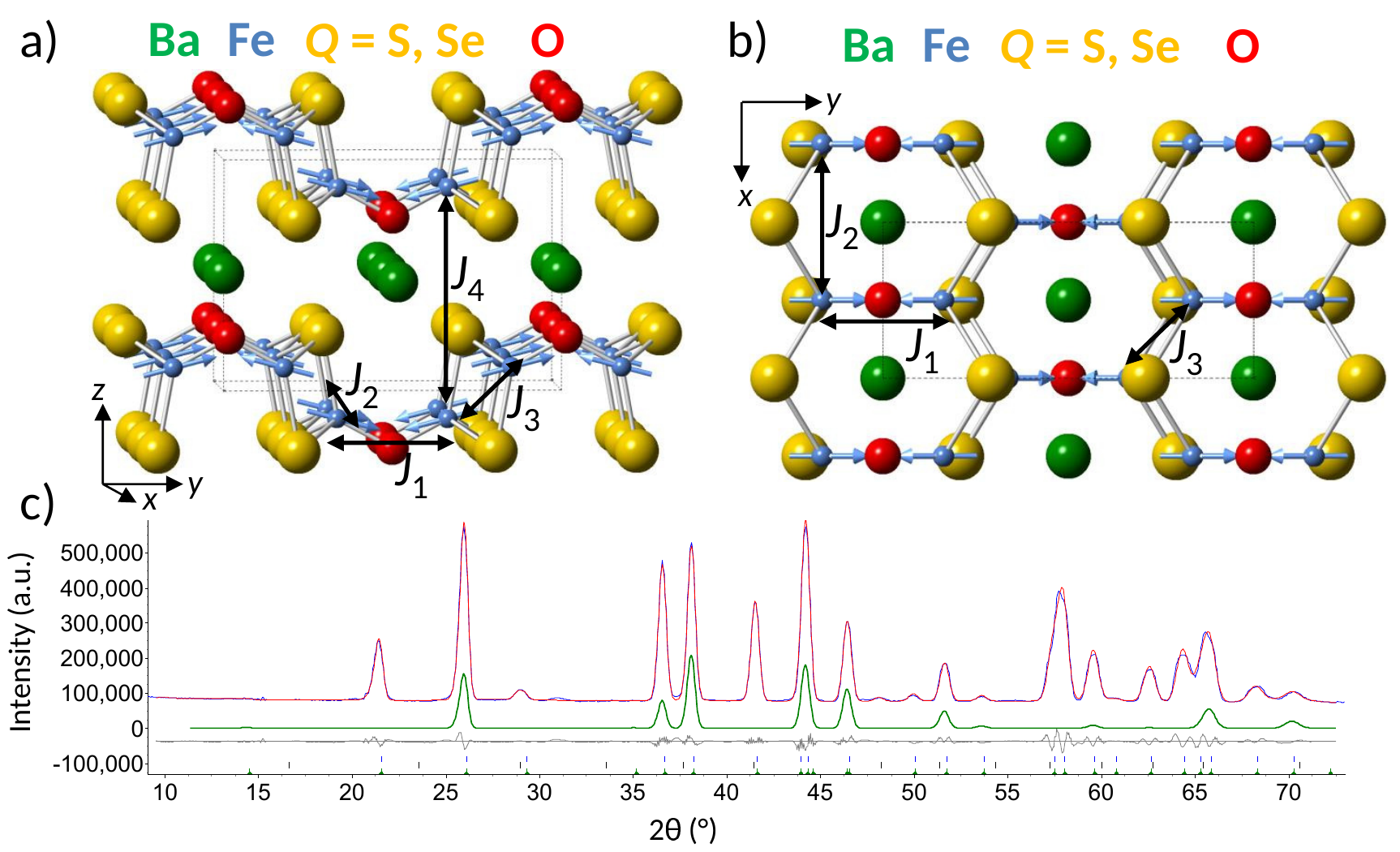}
\caption{[color online]  Illustration of $\Gamma1-$ magnetic structure showing Fe$^{2+}$ moments as blue arrows, viewed down a) [100] direction and b) [001] direction, and c) shows refinement profiles for BaFe$_{2}$S$_{2}$O using 2 K NPD data with observed, calculated and difference profiles are shown in blue, red and grey, respectively; upper blue ticks, middle black ticks and lower green ticks show reflection positions for BaFe$_{2}$S$_{2}$O, Fe$_{3}$O$_{4}$  and the magnetic phase, respectively, and scattering from the magnetic phase is highlighted in green.}
\label{BaFe2Q2O_mag_2K}
\end{figure}

No additional reflections were observed in low temperature NPD data, but the intensity of $0kl$ and $hk0$ reflections increased smoothly on cooling, whilst there was little change in $0k0$ reflections (see supplementary materials). These observations were consistent with long-range, three-dimensional magnetic order developing below $T_{\mathrm{N}}$ with $k$ vector $k$=(0 0 0). ISODISTORT~\cite{isodistort} was used to obtain descriptions of possible magnetic structures consistent with this $k$ vector. The collinear $\Gamma1-$ AFM structure (Figure \ref{BaFe2Q2O_mag_2K}), with moments oriented along [010] with AFM coupling across Fe - O - Fe rungs ($J_{1}$) and between ladders ($J_{3}$) but FM coupling along ladder legs ($J_{2}$), gave a good fit to the data and this model was used for subsequent analysis. This magnetic structure can be described by $Pm'm'n'$ symmetry and we note that this $\Gamma1-$ model also allows an AFM out-of-plane component. This cants the moments slightly away from the [010] direction to lie closer to the Fe - O bond direction, and including this additional parameter gave a very slight improvement in fit ($R_{wp}$ decreased from 3.807\% to 3.785\% for BaFe$_{2}$S$_{2}$O and from 3.392\% to 3.376\% for BaFe$_{2}$Se$_{2}$O). This allows a small AFM component of the moment along [001] and at 2 K, results in Fe$^{2+}$ moments canted at $\sim5^{\circ}$. Given the slight improvement in fit and the small refined component along [001], we cannot confirm this canting from our NPD data (a good fit is obtained with moments oriented along [010]) and no change in moment direction could be detected on cooling (see supplementary materials). However, this symmetry-allowed [001] component is compatible with fluctuations of the moments within the (011) planes that might eventually freeze out at low temperature, as discussed further below. Details from low temperature refinements are given in Table \ref{table_refine_2K}. The magnetic structure is illustrated in Figure \ref{BaFe2Q2O_mag_2K} with refinement profiles (see also supplementary materials).

\begin{table}[ht]
\caption{Refinement details and selected distances, bond lengths and angles from Rietveld refinements using 1.8 K NPD data for BaFe$_{2}$S$_{2}$O and 2 K NPD data for BaFe$_{2}$Se$_{2}$O using $Pmmn$ nuclear model and $\Gamma1-$ magnetic model for both.}
\centering
\begin{tabular} {c c c}
\hline\hline
X & $Q$= S, 1.8 K &  $Q$= Se, 2 K\\
\hline\hline
$a$ (\AA)& 3.9975(1) & 4.12633(6) \\
$b$ (\AA) & 9.5460(5) & 9.8378(1) \\
$c$ (\AA) & 6.4528(2) & 6.7006(1) \\
volume (\AA $^3$) & 246.24(2) & 272.004(8) \\
Ba 2$a$ $z$ & 0.5319(8) & 0.5111(2) \\
Ba $U_{\mathrm{iso}}$ $\times$ 100 (\AA $^2$) & 1.0(3) & 0.25(5) \\
Fe 4$e$ $y$ & 0.6657(3) & 0.66391(6) \\
Fe 4$e$ $z$ & 0.8823(3) & 0.88159(8) \\
Fe $U_{\mathrm{iso}}$ $\times$ 100 (\AA $^2$) & 1.0(3) & 0.80(3) \\
Fe moment ($\mu_{B}$) & 3.17(3) & 3.39(1) \\
Fe moment along [010] ($\mu_{B}$) & 3.16(3) & 3.38(1) \\
Fe moment along [001] ($\mu_{B}$) & 0.25(6) & 0.33(2) \\
Se 4$e$ $y$ & 0.7831(9) & 0.79209(6) \\
Se 4$e$ $z$ & 0.757(1) & 0.75776(9) \\
Se $U_{\mathrm{iso}}$ $\times$ 100 (\AA $^2$) & 1.0(3) & 0.63(4) \\
O 2$b$ $z$ & 0.7316(8) & 0.7380(1) \\
O $U_{\mathrm{iso}}$ $\times$ 100 (\AA $^2$) & 1.0(3) & 0.59(5) \\
$R_{wp}$  (\%) & 3.78 & 3.38 \\
$R_{p}$  (\%) & 2.84 & 3.99 \\
$\chi^{2}$  & 33.32 & 12.12 \\
Fe - Fe [010] (\AA) & 3.163(5) & 3.225(1) \\
Fe - Fe [111] (\AA) & 2.982(3) & 3.1053(7) \\
Fe - O (\AA) & 1.857(3) & 1.8779(8) \\
Fe - $Q$ [001] (\AA) & 2.379(7) & 2.4550(8) \\
Fe - $Q$ [110] (\AA) & 2.430(5) & 2.5564(4) \\
Fe - O - Fe ($^{\circ}$) & 116.8(3) & 118.35(7) \\
Fe - $Q$ - Fe [100] ($^{\circ}$) & 111.6(3) & 107.62(2) \\
Fe - $Q$ - Fe [111] ($^{\circ}$) & 76.6(2) & 76.55(2) \\
\hline
\label{table_refine_2K}
\end{tabular}
\end{table}

\subsection{Variable temperature NPD analysis}
\begin{figure}[t]
\includegraphics[width=0.95\linewidth]{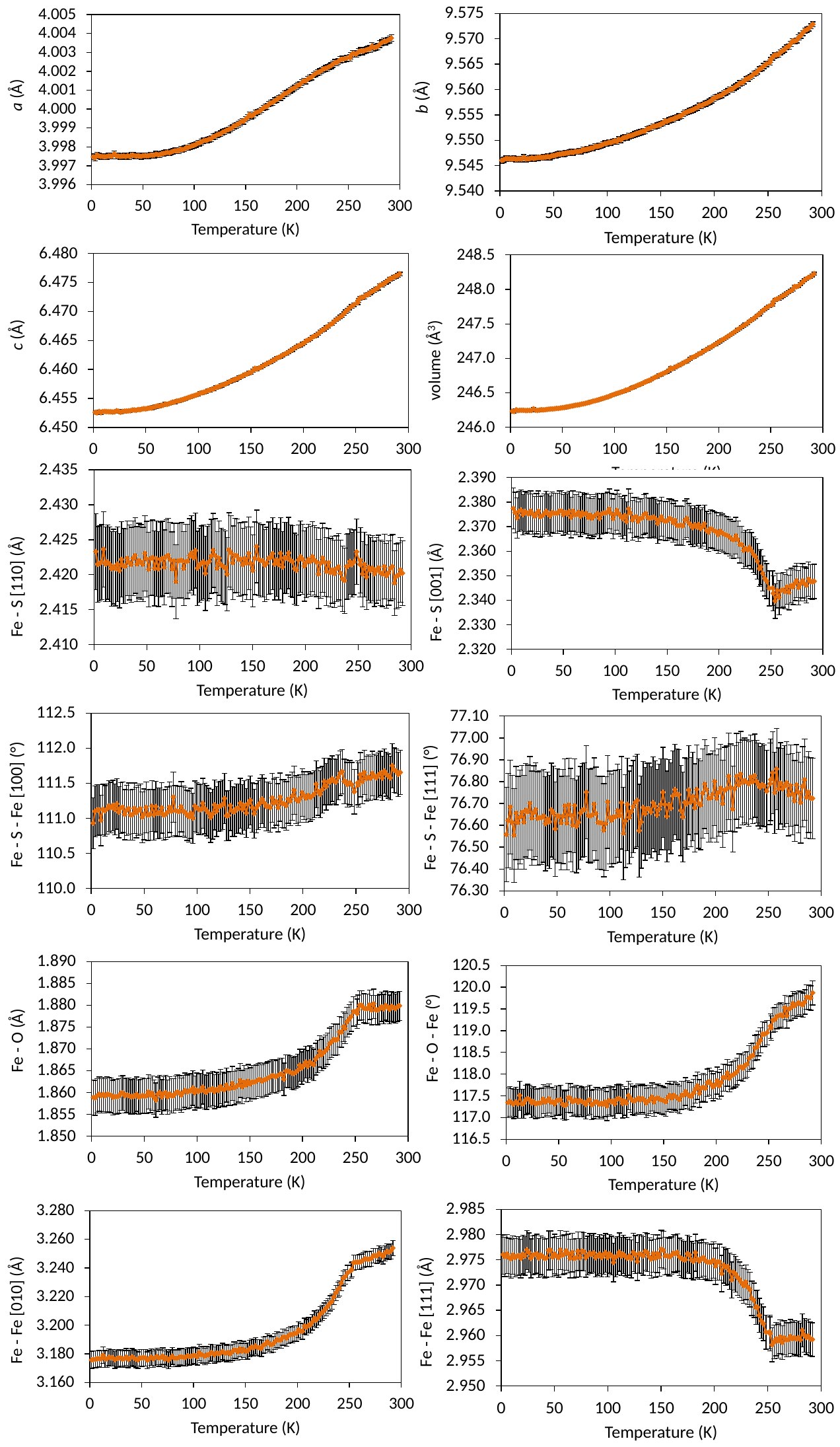}
\caption{[color online]  Unit cell parameters and selected distances, bond angles and lengths for BaFe$_{2}$S$_{2}$O from sequential refinements using variable temperature NPD data.}
\label{BaFe2Q2O_VT}
\end{figure}

\begin{figure}[t]
\includegraphics[width=0.95\linewidth]{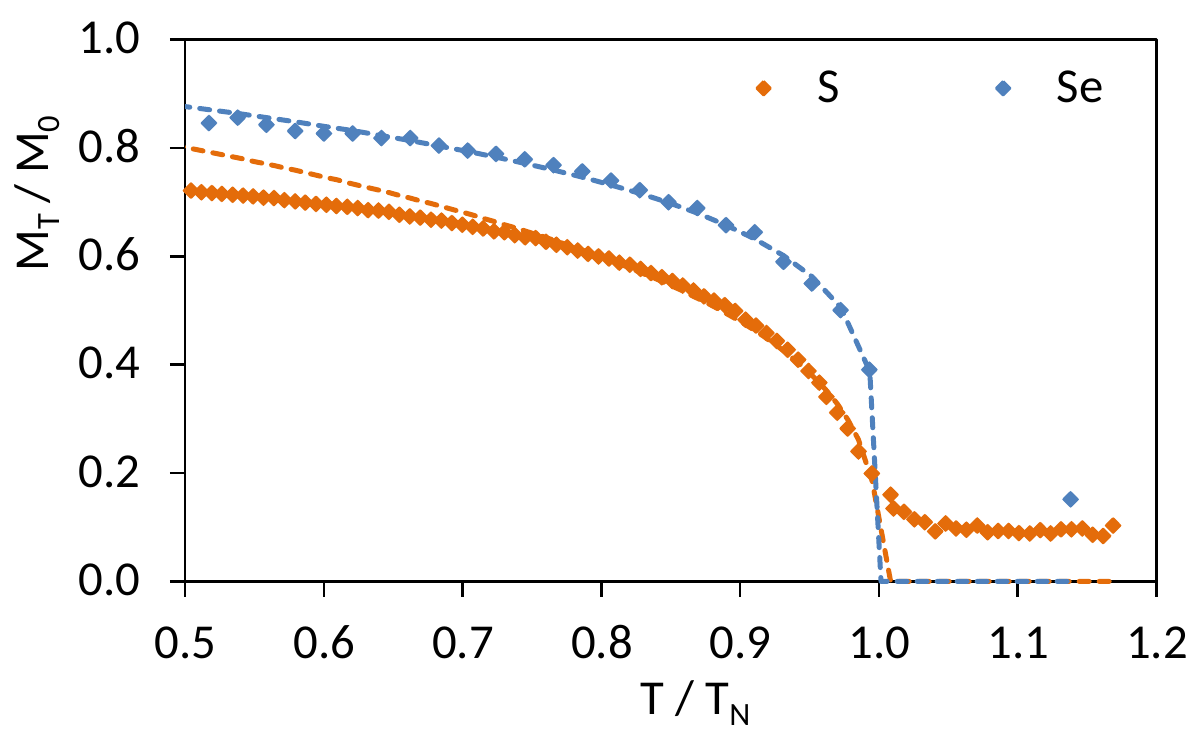}
\caption{[color online]  Evolution of magnetic order for BaFe$_{2}Q_{2}$O; data points are filled and the dashed lines are guides to the eye showing critical behaviour $M_{T}=M_\mathrm{0}[1-(\frac{T}{T_{N}})]^{\beta}$; critical exponent $\beta = 0.319(6)$, $T_{\mathrm{N}}$ = 249.7(1) K and $M_{\mathrm{0}}$ = 4.22(4) $\mu_{B}$ for BaFe$_{2}$S$_{2}$O; critical exponent $\beta = 0.190(5)$, $T_{\mathrm{N}}$ = 241.6(2) K and $M_{\mathrm{0}}$ = 3.95(3) $\mu_{B}$ for BaFe$_{2}$Se$_{2}$O.}
\label{critical_mag}
\end{figure}

Sequential Rietveld refinements were carried out on variable temperature NPD data and indicated that the unit cell parameters for both phases decrease smoothly on cooling (Figure \ref{BaFe2Q2O_VT} and supplementary material). No additional reflections were observed and there was no evidence to suggest changes to the long-range crystal structure or symmetry on cooling. Possible structural distortions were considered but did not give improvements in fit. We cannot rule out the possibility that a short-range Peierls-like distortion as observed for BaFe$_{2}$Se$_{3}$~\cite{Caron} may occur as our analysis of the long-range, average structure will not be sensitive to this. For BaFe$_{2}$S$_{2}$O, the Fe - O and second nearest-neighbour Fe - Fe distance (along [010]) decrease more abruptly below $T_{\mathrm{N}}$, also decreasing the Fe - O - Fe angle. The Fe - S - Fe angles change very little on cooling, whilst the Fe - S [001] bond length (bridging between the ladders) increases below $T_{\mathrm{N}}$, as does the interladder Fe - Fe distance (labelled Fe - Fe [111]), Figure \ref{BaFe2Q2O_VT}. This is in contrast to the selenide analogue,  BaFe$_{2}$Se$_{2}$O, for which all bond lengths and Fe - Fe distances decrease smoothly on cooling (supplementary material). The evolution of the magnetic order in  BaFe$_{2}Q_{2}$O can be fitted to critical behaviour (Figure \ref{critical_mag}), with critical exponents of 0.319(6) for  BaFe$_{2}$S$_{2}$O and 0.190(5) for  BaFe$_{2}$Se$_{2}$O.

\subsection{Muon spin relaxation analysis for BaFe$_{2}$Se$_{2}$O}
Muon spin relaxation ($\mu$SR) asymmetry data collected in zero applied field (ZF) were carried out at several temperatures to characterise the behaviour of the material across the transitions identified through the magnetisation measurements. Magnetisation and NPD have shown similar magnetic behaviour between the two systems and hence we anticipate the additional understanding provided by $\mu$SR measurements is likely to apply to both.
The evolution with time of the asymmetry for BaFe$_{2}$Se$_{2}$O was fitted using a simple exponential decay plus a constant background (Equation \ref{muons}).  
\begin{eqnarray}
A(t)= A_{0}e^{-\lambda t} + A_{back}
\label{muons}
\end{eqnarray}
\noindent where $A_{back}$ is the flat background, $A_{0}$ is the initial asymmetry, $\lambda$ the relaxation constant and $t$ is time. The background constant was fitted for the temperature region where the depolarisation rate is the fastest (around 150 K), where it took a value of 0.11$\pm$0.01. This was then fixed as the value of $A_{back}$ for the fits at every other temperature. The higher than usual value is due to the fact that the sample was loaded in several individual pouches of silver foil. 

\begin{figure}[t]
\includegraphics[width=0.95\linewidth]{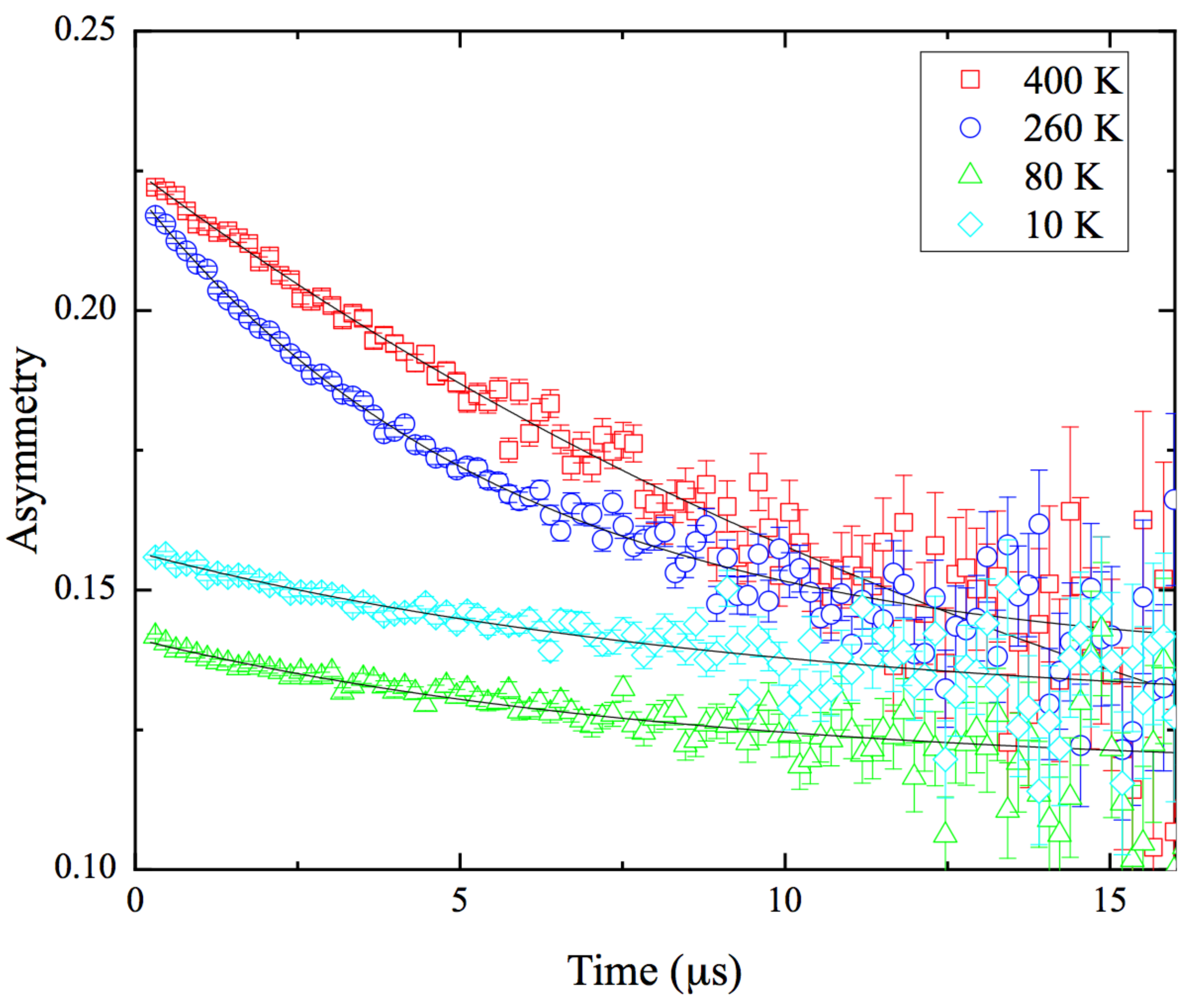}
\caption{[color online] {ZF asymmetry data for BaFe$_{2}$Se$_{2}$O at 400 K, 260 K, 80 K and 10 K. The four data sets illustrate the different magnetic environments encountered by the muons as a function of temperature as well as the quality of the fits to the model presented in Equ.\ref{muons}.}}
\label{muons_raw}
\end{figure}

\begin{figure}[t]
\includegraphics[width=0.95\linewidth]{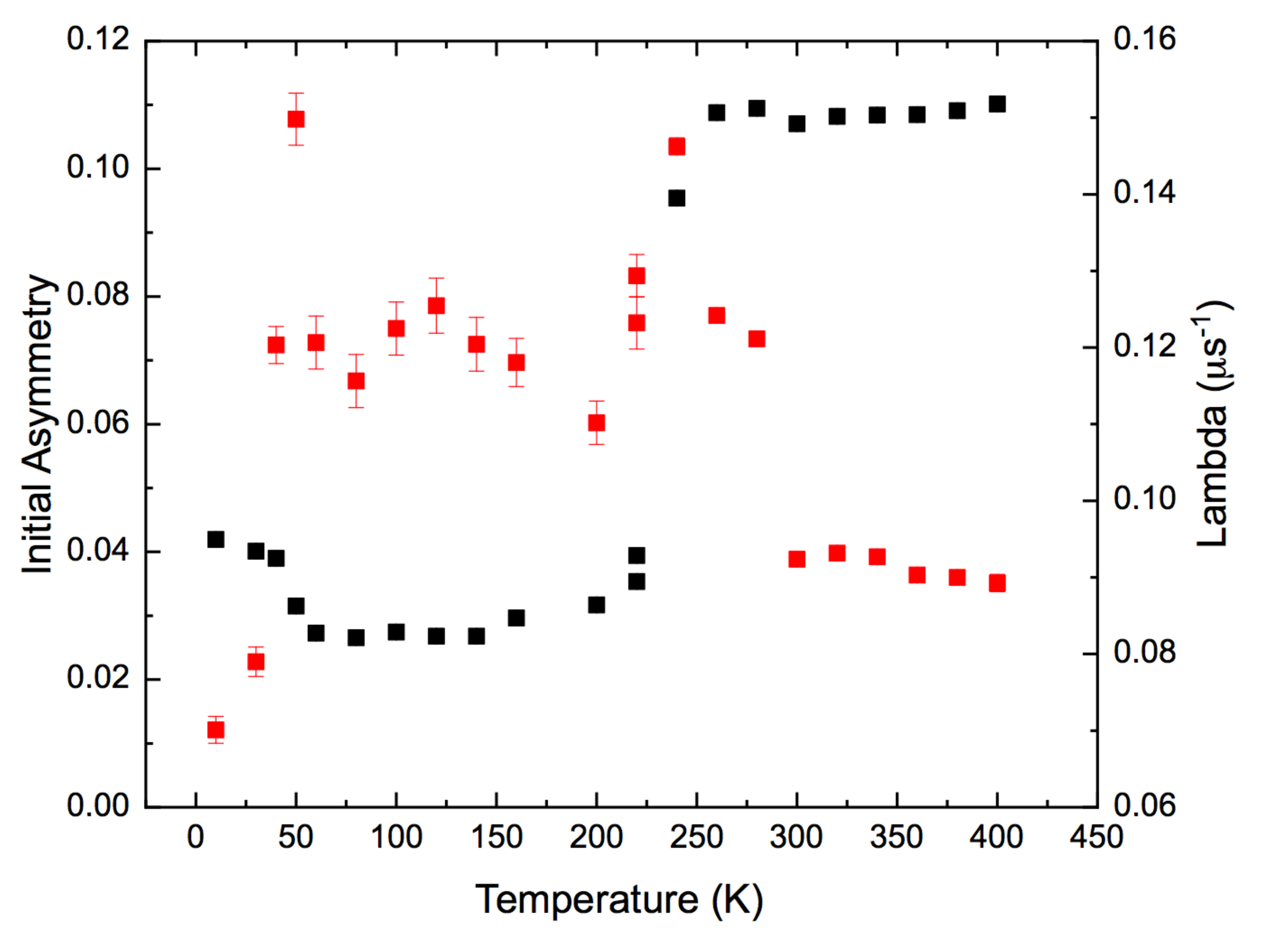}
\caption{[color online] {Evolution of the initial asymmetry (in black) and relaxation rate, $\lambda$,(in red) as a function of temperature. Both fitting parameters show the presence of two transitions (at $\sim$ 240 K and $\sim$ 40 K). Note that the transition at $\sim$ 240K has a broader signature in temperature than that at $\sim$ 50K.}}
\label{muons_fit}
\end{figure}

The four sets of raw data shown in Figure \ref{muons_raw} are representative of the different types of magnetic behaviour observed in this material as a function of temperature. Figure \ref{muons_fit} shows the values of the two fitting parameters at all measured temperatures and allow us to identify two transitions. The first one, at $\sim$240 K, is consistent with the AFM ordering transition at $T_{\mathrm{N}}$, observed also in the magnetisation measurements and by temperature-dependent neutron diffraction. The signature of the transition in the relaxation constant is noticeably broader than might have been expected for a typical three-dimensional magnetic phase transition and this may reflect the quasi-2D nature of the magnetic correlations above $T_{\mathrm{N}}$ observed in magnetic susceptibility data and emphasised by Han \textit{et al.}~\cite{Han}. In addition, the zero-field (ZF) muon relaxation data show the presence of a second transition at a lower temperature, around 40-50 K, consistent with the low temperature transition $T_{\mathrm{2}}$ observed in magnetic susceptibility measurements (see supplementary material and Figure \ref{BaFe2Q2O_structure_susceptibility}). As discussed above, NPD data are not sensitive to this magnetic transition, suggesting that it may correspond to a dynamic process. $\mu$SR measurements are sensitive to spin fluctuations with time scales typically in the range between 10$^{-12}$ and 10$^{-5}$ s and can detect slower dynamics than neutron scattering. A possible interpretation of this result is that the magnetically ordered state at 240 K contains moments whose canting is fluctuating at a rate for which neutron diffraction would only be sensitive to the average structure. This low temperature transition below 40 K may involve freezing out of the spin fluctuations within the (011) planes (which are allowed by the model used to fit the diffraction data) to give ordered components along both [010] and [001] below $\sim$50 K, resulting in an increase of the initial asymmetry as seen by the muons. 

\subsection{Electronic structure calculations}
\begin{figure}[t]
\includegraphics[width=0.95\linewidth]{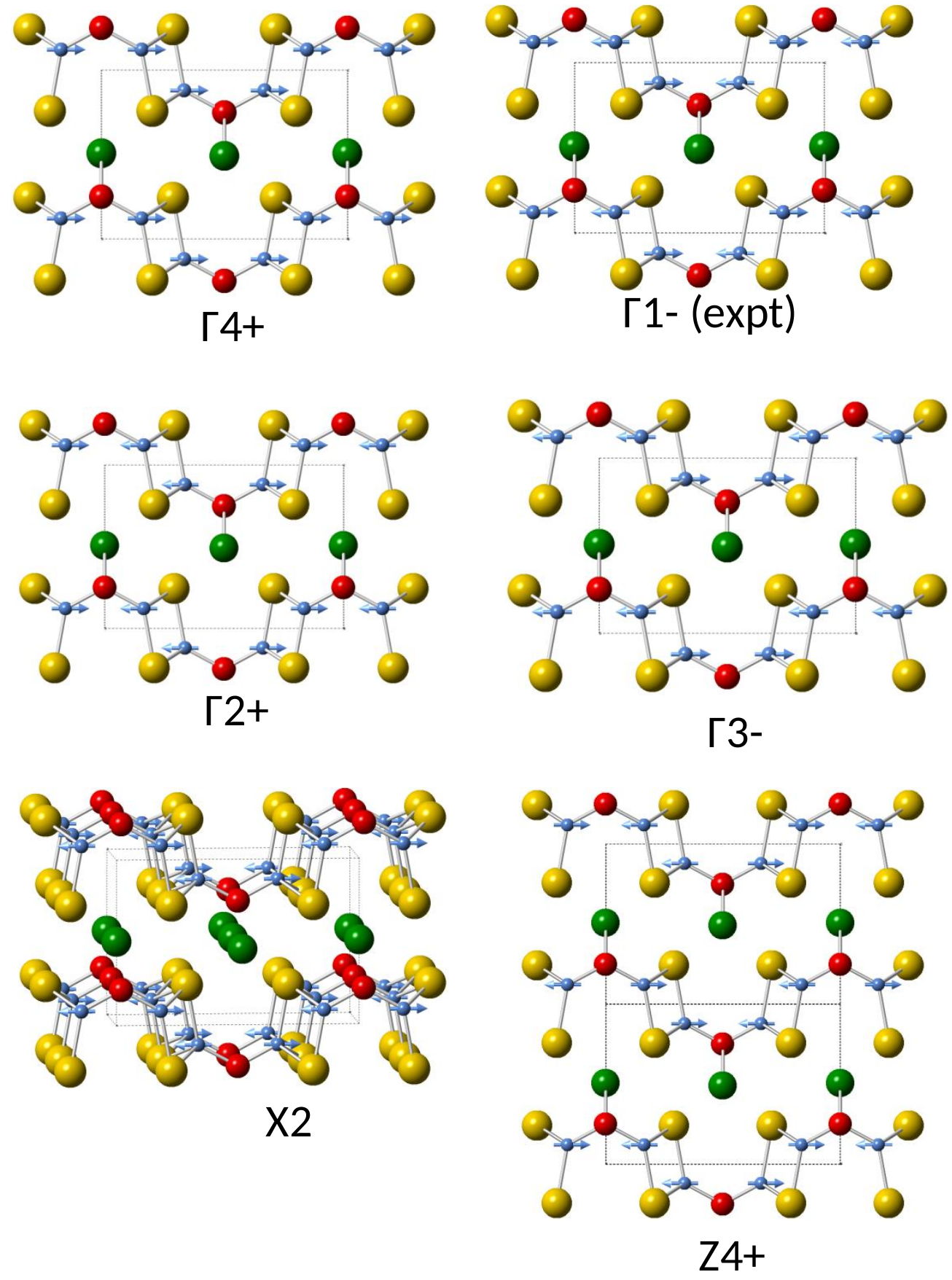}
\caption{[color online]  Five ordered spin states FM ($\Gamma4+$) and AFM $\Gamma1-$ (observed experimentally), $\Gamma2+$, $\Gamma3-$, X2 and Z4+ used to determine exchange interactions $J_{1}$ - $J_{4}$ for BaFe$_{2}Q_{2}$O, with barium, iron, chalcogenide and oxide ions shown in green, blue, yellow and red, respectively, and Fe$^{2+}$ spins shown by blue arrows.}
\label{DFT_states}
\end{figure}

We performed electronic structure calculations in order to estimate the exchange interactions for the material. Figure \ref{BaFe2Q2O_mag_2K} illustrates the four spin interactions we investigate in BaFe$_{2}Q_{2}$O: $J_{1}$ Fe - O - Fe ($\sim120^{\circ}$) across the "rungs" of the ladder, $J_{2}$ Fe - $Q$ - Fe ($\sim110^{\circ}$) along the "legs" of the ladder, $J_{3}$ "interladder" Fe - $Q$ - Fe ($\sim77^{\circ}$) interactions and interlayer $J_{4}$ interactions. To determine theoretical values of these exchange interactions, six ordered spin states (FM ($\Gamma4+$), and five AFM states), presented in Figure \ref{DFT_states}, were considered. $\Gamma1-$ (the model suggested by refinement of diffraction data) and $\Gamma2+$ both contain AFM $J_{1}$ (across the ladder rungs) and FM $J_{2}$ (along the ladder lengths), but differ in the sign of the interladder $J_{3}$ exchange, with AFM $J_{3}$ for $\Gamma1-$ and FM $J_{3}$ for $\Gamma2+$. The $\Gamma3-$ model is similar to the $\Gamma1-$ model with FM $J_{2}$ (along the ladder lengths) and AFM interladder $J_{3}$, but differs with FM $J_{1}$ (across the ladder rungs). The X2 model is analogous to the $\Gamma1-$ model but with AFM $J_{2}$ (along the ladder lengths), requiring doubling of the magnetic unit cell along [100]. These five models allow the three intralayer exchange interactions to be determined, but the weaker interlayer $J_{4}$ interactions require a magnetic unit cell doubled along [001] and so the Z4+ model, with AFM $J_{4}$ was also considered. 

The experimentally observed magnetic structure, $\Gamma1-$, is the calculated ground state for the range of on-site Coulomb repulsion $U_{Fe}$ studied (see Table \ref{relative_energies}). The $\Gamma1-$ and $\Gamma2+$ models differ only in the sign of the interladder $J_{3}$ interaction and the large difference in energies of these two models clearly indicates strong and AFM $J_{3}$ interactions. Likewise, the $\Gamma1-$ and $\Gamma3-$ models differ in the sign of $J_{1}$ exchange and again, the relative energies of these models indicate that $J_{1}$ interactions are stronger still and AFM. The AFM model X2, with AFM $J_{1}$ and AFM $J_{2}$ (in which half the $J_{3}$ interactions are FM and half are AFM) is stabilised with respect to the FM model (and also the $\Gamma2+$ and $\Gamma3-$ models with FM $J_{3}$ and $J_{1}$, respectively), but not to the extent of the $\Gamma1-$ model with FM $J_{2}$, suggesting that the $J_{2}$ exchange along the ladders is much weaker than the $J_{1}$ and $J_{3}$ exchange interactions. It is interesting that the FM model is found to be metallic for lower values of $U_{Fe}$ ($U_{Fe}$ = 2.0, 3.5 and 5.0 eV for $Q$=S and for $U_{Fe}$ = 2.0, 3.5 for $Q$=Se), consistent with theoretical calculations reported by Han \textit{et al.}~\cite{Han}.

The total spin exchange energies of these magnetic models can be expressed in terms of the spin Hamiltonian,
\begin{eqnarray}
H=-\sum_{i,j} J_{i,j}\vec{S}_{i}\cdot\vec{S}_{j}.
\label{hamiltonian}
\end{eqnarray}
\noindent where $J_{i,j}$  is the spin exchange interaction between the spin sites $i$ and $j$ and can take values $J_{1}$ to $J_{4}$, as appropriate. By applying the energy expression for spin dimers with $N$ unpaired spins per spin site (four for Fe$^{2+}$), the total spin energies per unit cell can be written as 
\begin{eqnarray}
E = (n_{1}J_{1} + n_{2}J_{2} + n_{3}J_{3} + n_{4}J_{4}) \left({N_{Fe}^{2}\over 4}\right) 
\label{energy}
\end{eqnarray}
\noindent where $N_{Fe}$ = 4, and the coefficients $n_{1}$ to $n_{4}$ for the five spin ordered states are given in supplementary material.

\begin{table}[ht]
\caption{Energies (in meV per unit cell) of AF spin arrangements shown in Figure \ref{DFT_states} relative to $\Gamma1-$ AFM arrangement for various $U_{Fe}$ values (in eV) for BaFe$_{2}Q_{2}$O. $^{\dag}$ Values for these doubled magnetic cells were halved for comparison with the other $k$=(0 0 0) magnetic models. Where data is missing, calculations were either not performed, or the magnetic ordering produced a metallic ground state.}
\centering
\begin{tabular} {c c c c c c c}
\hline\hline
$U_{Fe}$ & $Q$  & $\Gamma2+$ & $\Gamma4+$ & $\Gamma3-$ & X2$^{\dag}$ & Z4+$^{\dag}$ \\
\hline\hline
2.0 & S &  - & - & 401 & - & - \\
3.5 & S &  778 & - & 309 & 114 & 776 \\
5.0 & S &  689 & 939 & 234 & 102 & 687 \\
6.5 & S &  597 & 783 & 175 & 86 & 596 \\
\hline
2.0 & Se &  - & - & 369 & 95 & 659 \\
3.5 & Se &  - & - & 279 & 101 & 671 \\
5.0 & Se &  583 & 792 & 208 & 87 & 580 \\
6.5 & Se &  495 & 649 & 153 & 71 & 493 \\
\hline
\label{relative_energies}
\end{tabular}
\end{table}

The relative energies of these magnetic models can be used, with energies for the models calculated from Equation \ref{energy}, to determine theoretical values of the exchange interactions, as shown in Table \ref{table_energy_mapping}~\cite{Whangbo1,Whangbo2,Whangbo3,Whangbo4,Whangbo5,Ce2O2FeSe2}. In contrast to the theoretical work by Han \textit{et al.}~\cite{Han}, this analysis suggests that the relative strength of the Fe-O-Fe $J_{1}$ interaction, compared with the Fe - Q - Fe $J_{2}$ interaction increases with $U$.

Our calculations suggest that longer-range exchange interactions (beyond next-nearest neighbour) are negligible as suggested previously by Han \textit{et al.}~\cite{Han}; this is supported by calculations for the doubled unit cells X2 and Z4+ whose total energies are consistent with only $J_{1-4}$ interactions. Indeed, the very low values calculated for $J_{4}$ (see Table \ref{table_energy_mapping}) equal zero within the uncertainty expected for these calculations ($\sim$ 0.1 meV). 

A value of $U_{Fe}$=3.5 eV appears to simultaneously minimise both the maximum force on any species and the stress on the lattice, suggesting $U_{Fe}$=3.5 eV would likely produce a relaxed structure closest to experiment (see supplementary materials). This value of $U_{Fe}$ also produces a magnetic moment close to experiment (for BaFe$_{2}$Se$_{2}$O, $\mu_{Fe}$ = 3.45 $\mu_{B}$ from calculations compared with 3.39(1) $\mu_{B}$ from experiment, see Table \ref{table_refine_2K}), and hence might be a good estimate of the on-site Coulomb repulsion of Fe in this environment.

\begin{table}[ht]
\caption{Values of $J_{1}$, $J_{2}$, $J_{3}$ and $J_{4}$ (in meV) from energy-mapping analyses based on various $U_{Fe}$  values (in eV) for BaFe$_{2}Q_{2}$O.}
\centering
\begin{tabular} {c c c c c c}
\hline\hline
$U_{Fe}$ & $Q$  & $J_{1}$ & $J_{2}$ & $J_{3}$ & $J_{4}$\\
\hline\hline
2.0 & S & -25.1 & - & - & - \\
3.5 & S & -19.3 & -8.6 & -24.3 & -0.1 \\
5.0 & S & -14.6 & -7.6 & -21.5 & -0.1 \\
6.5 & S & -10.9 & -6.6 & -18.7 & -0.1 \\
\hline
2.0 & Se & -23.1 & - & - & - \\
3.5 & Se & -17.5 & - & - & -\\
5.0 & Se & -13.0 & -6.4 & -18.2 & -0.1 \\
6.5 & Se & -9.6 & -5.5 & -15.5 & -0.1 \\
\hline
\label{table_energy_mapping}
\end{tabular}
\end{table}

\section{Discussion}
Analysis of our powder diffraction and magnetic susceptibility results for BaFe$_{2}Q_{2}$O are consistent with experimental results reported by Valldor and Huh~\cite{Valldor, Huh} which illustrate the difficulty in preparing samples free from Fe$_{3}$O$_{4}$ and Fe$Q$ impurities. The traces of Fe$_{3}$O$_{4}$ (demonstrated unambiguously for our samples by combined NPD and magnetic susceptibility data) make it likely that the magnetic phase transition at $\sim$115 K reported by Lei \textit{et al.} ~\cite{Lei} (and observed in our low magnetic field susceptibility data, see supplementary material) is likely the Verwey transition in Fe$_{3}$O$_{4}$~\cite{Verwey}. To the best of our knowledge, this is the first time this 115 K transition in BaFe$_{2}Q_{2}$O has been attributed to this magnetic impurity phase.

The NPD data presented here also give information on the long-range, average crystal structure and, within the sensitivity of the refinements, there is no evidence for antisite disorder (although we cannot rule out this possibility at a more local scale as suggested by Lei \textit{et al.} ~\cite{Lei} from M\"{o}ssbauer results). Similarly, Popovic \textit{et al.} ~\cite{Popovic} suggest that a structural change may occur on cooling below $T_{\mathrm{N}}$ based on Raman spectroscopy data for BaFe$_{2}$Se$_{2}$O and although our NPD data give no evidence for a change in symmetry, there is a clear change in the iron coordination environment at $T_{\mathrm{N}}$ with contraction of Fe-Fe distances (along the ladder rungs) as the Fe-O and Fe-O-Fe bond lengths and angles both decrease (see Figure \ref{BaFe2Q2O_VT} and supplementary material). In terms of interladder distances (Fe-$Q$ [001] and Fe-Fe [111]), whilst these contract slightly on cooling for BaFe$_{2}$Se$_{2}$O (supplementary material), surprisingly, both distances increase noticeably below $T_{\mathrm{N}}$ for BaFe$_{2}$S$_{2}$O.

These changes in Fe$^{2+}$ coordination environment at $T_{\mathrm{N}}$ are likely coupled to the magnetic ordering and the increasing interladder distance might be expected to weaken the interladder exchange $J_{3}$. We note that this AFM interaction competes with the much weaker AFM $J_{2}$ exchange along the lengths of the ladders but with a similar degree of frustration for both BaFe$_{2}$Se$_{2}$O and BaFe$_{2}$S$_{2}$O (see further discussion below), it is unclear why this structural change occurs (which is likely to weaken $J_{3}$) in the sulfide analogue and not the selenide.

A short-range spin-Peierls like distortion was observed from n-PDF analysis for the 2-leg ladder system BaFe$_{2}$Se$_{3}$~\cite{Caron}. These 2-leg ladder systems contain double chains of Fe$Q_{4}$ tetrahedra and have more itinerant electronic structures ~\cite{Takubo}. It is unlikely that such distortions would occur in the more localised BaFe$_{2}Q_{2}$O materials ($U_{Fe}\sim$ 3.5 eV). In addition, our time-of-flight NPD data for BaFe$_{2}$Se$_{2}$O (see supplementary material) might have been expected to show diffuse scattering if local distortions occurred. However, we cannot rule out the presence of local distortions, as our analysis of the long-range, average crystal structure has limited sensitivity to them.

The experimentally observed magnetic structure for BaFe$_{2}Q_{2}$O (Figure \ref{BaFe2Q2O_mag_2K}) is similar to that proposed by Han \textit{et al.} and consistent with magnetisation measurements on single crystals, which suggested that the easy axis of magnetisation is within the $ab$ plane~\cite{Han}. This structure is also in agreement with M\"{o}ssbauer studies on BaFe$_{2}$S$_{2}$O and on SrFe$_{2}Q_{2}$O ($Q$ = S, Se) by Huh and Valldor \textit{et al.}~\cite{Huh,Valldor}, which indicated a simple, collinear AFM structure for these systems, and similar to the magnetic structure reported by Guo \textit{et al.} for SrFe$_{2}Q_{2}$O ~\cite{Guo}. The magnitude of the magnetic moments determined from NPD refinements (3.15(3) $\mu_{B}$ and 3.31(1) $\mu_{B}$ for BaFe$_{2}$S$_{2}$O and BaFe$_{2}$Se$_{2}$O, respectively, Table \ref{table_refine_2K}) are similar to those reported for other insulating iron oxychalcogenides (e.g. 3.14(8) $\mu_{B}$ for Ce$_{2}$O$_{2}$FeSe$_{2}$~\cite{Ce2O2FeSe2}, 3.50(2) $\mu_{B}$ for La$_{2}$O$_{2}$Fe$_{2}$OSe$_{2}$~\cite{La2O2Fe2OSe2}) and in the parent phase to superconducting K$_{0.8}$Fe$_{1.6}$Se$_{2}$ (3.31 $\mu_{B}$)~\cite{KxFe2-ySe2}. 

This $\Gamma1-$ magnetic structure can be understood in terms of the dominant AFM exchange interactions $J_{3}$ and $J_{1}$. The comparable strength of these exchange interactions (Table \ref{table_energy_mapping}) brings into question earlier descriptions of these BaFe$_{2}Q_{2}$O systems as "spin ladders"~\cite{Popovic,Valldor,Huh,Guo} because the "interladder" exchange $J_{3}$ is comparable or stronger than $J_{1}$ (across the ladder "rungs") depending on $U_{Fe}$. The weak FM coupling along the lengths of the ladders results from two strong AFM $J_{3}$ interactions between the Fe$^{2+}$ sites in adjacent ladders, frustrating the weaker AFM $J_{2}$ exchange between Fe$^{2+}$ sites along the length of the ladder. We note that the energy of a single $J_{2}$ interaction is comparable to the energy of the low temperature feature in magnetic susceptibility data and observed in muon spin relaxation experiments ($T_{\mathrm{2}}\sim$59 K for BaFe$_{2}$S$_{2}$O, $T_{\mathrm{2}}\sim$40 K for BaFe$_{2}$Se$_{2}$O). This frustration between $J_{2}$ and $J_{3}$ along the lengths of the ladder may result in some local/dynamic disorder for $T_{\mathrm{2}}\sim$ $<$ $T$ $<$ $T_{\mathrm{N}}$ that freezes out at low temperatures below $T_{\mathrm{2}}$ when $J_{2}$ becomes comparable to $k_{\mathrm{B}}T$. NPD data showed no evidence for any diffuse magnetic scatter that might arise from such disorder, but may not be sensitive to this if the fluctuations are very small, or are slower than the neutron timescale ($\sim10^{-13}$ s). It is likely that such spin fluctuations exist below $T_{\mathrm{N}}$ (with moments on average along [010]) before freezing out (to give the small AFM-ordered [001] component, Figure \ref{BaFe2Q2O_mag_2K}) $T_{\mathrm{2}}$ as observed in $\mu$SR data.  

The presence of frustration has been considered by Huh and by Valldor \textit{et al.} in the strontium analogues ~\cite{Huh,Valldor}. It is interesting that the results of our DFT calculations suggest a very similar degree of frustration (in terms of the relative magnitudes of $J_{2}$ and $J_{3}$, Table \ref{table_energy_mapping}) for these two barium analogues. Huh and Valldor \textit{et al.}~\cite{Huh,Valldor} note the higher degree of frustration for the strontium analogues SrFe$_{2}Q_{2}$O and suggest that this might relate to the size of the Fe-$Q$-Fe angles; our DFT results indicate that the degree of frustration relates to the relative magnitude of $J_{2}$ and $J_{3}$ which will be very sensitive to the Fe-$Q$-Fe angles, consistent with their hypothesis. Further calculations on the strontium analogues would be of interest to confirm this.

The analysis of the dimensionality of the magnetic order does not give a conclusive answer. Our DFT calculations show that the in-plane exchange interactions $J_{1}$, $J_{2}$ and $J_{3}$ are noticeably stronger than the interlayer coupling $J_{4}$ (Table \ref{table_energy_mapping}) and so it is unsurprising that magnetic susceptibility (see Figure \ref{BaFe2Q2O_structure_susceptibility}, and supplementary materials) and heat capacity measurements~\cite{Han} suggest short-ranged, two-dimensional magnetic correlations above $T_{\mathrm{N}}$. However, given the relative high temperatures for $T_{\mathrm{N}}$ below which magnetic Bragg scattering is observed and that there is a clear drop in the asymmetry of the muon decay (suggesting three-dimensional magnetic order), it is surprising that such low values are calculated for $J_{4}$. We note that these observations are similar to those for BaFe$_{2}Q_{3}$, with evidence for short-ranged magnetic correlations above $T_{\mathrm{2}}$~\cite{Lei2,Caron,Seidov} and interladder interactions of the same order of magnitude as $J_{4}$ interactions calculated here~\cite{Sefat,Mourigal}. This may indicate deficiencies in our model, for example, the magnetic anisotropy of the Fe$^{2+}$ site has been shown to be significant in other iron oxychalcogenides~\cite{La2O2Fe2OSe2} but has been neglected in this current study. Further calculations including the effects of spin-orbit coupling would be of interest to investigate this further. It is worth noting that the critical exponents, $\beta$ (see Figure \ref{critical_mag}) are similar to those expected for 3D and 2D Ising systems for BaFe$_{2}$S$_{2}$O and BaFe$_{2}$Se$_{2}$O, respectively. Given the similar values for exchange interactions calculated by DFT for these two materials, it is not clear why there should be a significant difference in their magnetic dimensionality. This question remains open and it is, in our opinion, worthy of further study both by experiments and computationally.

During preparation of this manuscript, we became aware of the magnetic structure reported for SrFe$_{2}Q_{2}$O~\cite{Guo} which is consistent with our findings here for BaFe$_{2}Q_{2}$O, suggesting that our conclusions are likely to apply to a wide range of materials in this structural family. 

\section{Conclusions}
Our results suggest that the magnetic insulators BaFe$_{2}Q_{2}$O differ from spin-ladder systems such as the superconducting BaFe$_{2}Q_{3}$ materials because the nearest-neighbour exchange interactions via oxide and selenide anions are of comparable strength, giving rise to stronger magnetic coupling. The more localised electronic structure with narrower Fe 3d bands (due to the harder oxide in the Fe$^{2+}$ coordination environment and the buckled Fe-$Q$-O layers~\cite{Han}) results in large ordered moments on the Fe$^{2+}$ site, in both the sulfide and selenide analogues, similar to other insulating oxychalcogenides. Although it would be interesting to investigate the extent to which the electronic structure of BaFe$_{2}Q_{2}$O materials could be tuned by electron-doping and applied pressure, the narrow Fe 3d bands (suggested by Han \textit{et al.}~\cite{Han} and consistent with our theoretical and experimental work) are likely to place the BaFe$_{2}Q_{2}$O materials further into the Mott insulating side than other Fe$^{2+}$ spin-ladder systems. 

\begin{acknowledgments}
EEM, SR and BDC are grateful to Pascal Manuel, Dmitry Kalyavin for assistance with NPD experiments carried out at WISH (ISIS, Rutherford Appleton Laboratory, U.K., DOI: 10.5286/ISIS.E.79113963), to ISIS for provision of muon beamtime (DOI: 10.5286/ISIS.E.67771885) and Emma Suard for assistance with NPD experiments carried out at D20 (ILL, Grenoble, France). We are also grateful to both facilities for the provision of beam time to conduct neutron and muon measurements. BDC is grateful to EPSRC and University of Kent his studentship. NCB acknowledges the Royal Commission for the Exhibition of 1851 for a fellowship, computational resources at Imperial College London's high-performance computing facility, and also the UK Materials and Molecular Modelling Hub for computational resources (partially funded by the EPSRC project EP/P020194/1).
\end{acknowledgments}

\bibliography{150518}

\end{document}